\documentstyle[12pt,epsf,rotating]{article}

\catcode`@=11
\def\citer{\@ifnextchar [{\@tempswatrue\@citexr}{\@tempswafalse\@citexr[]}}
 
\def\@citexr[#1]#2{\if@filesw\immediate\write\@auxout{\string\citation{#2}}\fi
  \def\@citea{}\@cite{\@for\@citeb:=#2\do
    {\@citea\def\@citea{--\penalty\@m}\@ifundefined
       {b@\@citeb}{{\bf ?}\@warning
       {Citation `\@citeb' on page \thepage \space undefined}}%
\hbox{\csname b@\@citeb\endcsname}}}{#1}}
\catcode`@=12

\def\AAcd{
F_\Box & = & \frac{1}{S^2}\left\{ \pht 4 S +  8 S m_Q^2 C_{ab}
-2S(T+U) m_Q^4 (D_{abc}+D_{bac}+D_{acb}) \right. \\
& & \left. +(\rho_c+\rho_d) m_Q^2\left[\pht T_1 C_{ac} + U_1 C_{bc} + U_2 C_{ad}
+ T_2 C_{bd} - (TU-\rho_c\rho_d) m_Q^2 D_{acb}\right] \right\} \\
G_\Box & = & \frac{1}{S(TU-\rho_c\rho_d)}\left\{ \pht (T^2+\rho_c\rho_d) m_Q^2
\left[ \pht S C_{ab} + T_1 C_{ac} + T_2 C_{bd} - ST m_Q^2 D_{bac} \right]
\right. \\
& & \hspace{2.60cm} + (U^2+\rho_c\rho_d) m_Q^2 \left[ \pht S C_{ab} + U_1 C_{bc}
+ U_2 C_{ad} - SU m_Q^2 D_{abc} \right] \\
& & \hspace{2.60cm} - (T^2+U^2-2\rho_c\rho_d) (T+U) m_Q^2 C_{cd} \\
& & \hspace{2.60cm} \left. - 2(T+U)(TU-\rho_c\rho_d) m_Q^4
(D_{abc}+D_{bac}+D_{acb}) \pht \right\}
}

\def\AA{\AAcd}

\def\text#1{\mbox{#1}}
 
\newcommand{\idnty}{1\hspace{-0.85mm}\mbox{l}}
\newcommand{\lsim}{\raisebox{-0.13cm}{~\shortstack{$<$ \\[-0.07cm] $\sim$}}~}

\newcommand{\nn}{\noindent}
\newcommand{\non}{\nonumber}
\newcommand{\tb}{\mbox{tg$\beta$}}
\newcommand{\s}{\hat{s}}

\newcommand{\mq}{m_Q^2}

\newcommand{\pht}{\phantom{\frac{1}{1}}\!\!\!\!}

\begin{document}

\begin{titlepage}

\begin{flushright}
DESY 95--215 \\
December 1995 \\
hep-ph/9603205
\end{flushright}

\vspace{1cm}

\begin{center}
\baselineskip25pt

{\large\sc Pair Production of Neutral Higgs Particles \\
           in Gluon--Gluon Collisions}

\end{center}

\vspace{1cm}

\begin{center}
\baselineskip12pt

\def\thefootnote{\fnsymbol{footnote}}

{\sc T.~Plehn$^1$, M.~Spira$^{2}$\footnote{Address after Jan.1,1996:
                        TH Division, CERN, CH--1211 Geneva 23, Switzerland}
and P.~M.~Zerwas$^1$} \\ 
\vspace{1cm}

$^1$ Deutsches Elektronen--Synchrotron DESY, D--22603 Hamburg, FRG \\
\vspace{0.3cm}

$^2$ II.~Institut f\"ur Theoretische Physik\footnote{Supported by
Bundesministerium f\"ur Bildung und Forschung (BMBF), Bonn, under Contract
05 6 HH 93P (5), and by EU Program {\it Human Capital and Mobility}
through Network {\it Physics at High Energy Colliders} under Contract
CHRX--CT93--0357 (DG12 COMA).}, D--22761 Hamburg, FRG \\
\vspace{0.3cm}
 
\end{center}

\vspace{2cm}

\begin{abstract}
  \normalsize \nn Pair production processes of neutral Higgs particles
  will allow us to study the trilinear Higgs couplings at future
  high--energy colliders. Several mechanisms give rise to multi--Higgs
  final states in hadron interactions. In the present paper we
  investigate Higgs pair production in gluon--gluon collisions.  After
  recapitulating pair production in the Standard Model, the analysis
  of the cross sections is carried out in detail for the neutral Higgs
  particles in the minimal supersymmetric extension.
\end{abstract}

\end{titlepage}

\def\thefootnote{\arabic{footnote}}
\setcounter{footnote}{0}

\setcounter{page}{2}

\section{Introduction}
The reconstruction of the Higgs potential is an experimental {\it
  prima facie} task to establish the Higgs mechanism as the basic
mechanism for generating the masses of the fundamental particles. This
task requires the measurement of the self--couplings of the Higgs
particles.\smallskip

In the Standard Model (SM) \cite{0} the trilinear and the quartic
couplings of the physical Higgs particle $H$ are uniquely fixed if the
Higgs mass is known.\smallskip 

In the minimal supersymmetric extension of the Standard Model (MSSM),
a large variety of couplings exists among the members $h,H,A,H^\pm$ of
the Higgs quintet \cite{4}. [$h$ and $H$ are the light and heavy
CP--even Higgs bosons\footnote{Following standard notations we have
  taken care that no confusion will arise from using the same symbol
  for the SM and the heavy CP--even MSSM Higgs particle.}, $A$ is the
CP--odd (pseudoscalar) Higgs boson, and $H^\pm$ is the charged Higgs
pair.]  While general two--doublet models contain three mass
parameters and seven real self--couplings in CP conserving theories,
the Higgs self--couplings are fixed in terms of gauge couplings in the
MSSM, and the mass parameters can be expressed by the two vacuum
expectation values of the neutral Higgs fields and one of the physical
Higgs masses. Since the sum of the squares of the vacuum expectation
values is given by the $W$ mass, tg$\beta$, the ratio of the two
vacuum expectation values, and $M_A$, the mass of the CP--odd Higgs
boson $A$, are generally chosen as the free parameters of the MSSM.
The trilinear and quartic Higgs self--couplings are determined by
those two parameters.\smallskip

The measurement of the Higgs self--couplings will be a very difficult
task. In the Standard Model the production cross section for $HH$
Higgs pairs are small, similarly the continuum production of Higgs
pairs in the MSSM. Only if heavy MSSM Higgs bosons can decay into
pairs of light Higgs bosons, the associated Higgs self--couplings can
be determined fairly easily by measuring the decay branching ratios.
Several aspects of multi--Higgs production have been discussed in the
literature.  Most analyses treat $HH$ pair production in the Standard
Model only at the theoretical level \cite{1a,1}. Within the MSSM, the
search for the heavy Higgs boson $H$ has been simulated in the
$b\bar{b}$ decay mode at the LHC \cite{2}. $h \to AA$ events have been
searched at LEP \cite{3}; more general aspects of multi--Higgs
production in $e^+e^-$ collisions have only recently been analyzed
theoretically in Ref.\cite{4}.\smallskip

Several mechanisms give rise to the production of pairs of neutral
Higgs bosons in hadron collisions. Multi--Higgs final states can be
produced through Higgs--strahlung off $W$ bosons and through $WW$
fusion in proton--proton collisions. Bremsstrahlung of Higgs particles
off heavy quarks can also be exploited. In the present paper we
discuss the production of Higgs pairs in gluon--gluon collisions at
the LHC.  We have determined the cross sections for the continuum in
the Standard Model
\begin{equation}
pp \to gg \to HH 
\end{equation}
as well as for the continuum and resonance decays in the minimal
supersymmetric theory
\begin{equation} 
pp \to gg \to \Phi_i\Phi_j  \qquad \qquad \Phi_i=h,H,A
\end{equation}
restricting ourselves to neutral Higgs bosons in the present
report\footnote{The derivation of the SM cross sections has nicely
  been described in Ref.\cite{1}. We will briefly discuss this case
  again to exemplify the techniques for the simplest scenario of Higgs
  pair production, before generalizing the analysis for unequal masses
  and pseudoscalar Higgs particles in the MSSM.}. It can be
anticipated from the large number of low--$x$ gluons in high--energy
proton beams that the $gg$ channel is of particular interest in the
continuum for fairly low Higgs masses. In the MSSM the on--shell
production of heavy Higgs bosons with subsequent Higgs cascade decays
will eventually provide a copious source of light multi--Higgs final
states. The cross sections are affected, besides the normal
Higgs--boson couplings to gauge bosons and fermions \cite{5}, by the
trilinear Higgs couplings \cite{5a}:
\begin{eqnarray}
&SM\hphantom{SM}:& \lambda_{HHH} = \frac{3 M_H^2}{M_Z^2}
\label{eq:3} \\
&MSSM:& \lambda_{hhh} = 3\cos(2\alpha) \sin(\beta+\alpha)
+ \frac{3 \epsilon}{M_Z^2} \frac{\cos^3\alpha}{\sin \beta} \non \\
&& \lambda_{Hhh} =2 \sin(2\alpha) \sin(\beta+\alpha) 
                - \cos(2\alpha) \cos(\beta+\alpha)
+ \frac{3 \epsilon}{M_Z^2} \frac{\sin\alpha\cos^2\alpha}{\sin \beta} \non \\
&& \lambda_{HHh} =- 2 \sin(2\alpha) \cos(\beta+\alpha) 
                - \cos(2\alpha) \sin(\beta+\alpha)
+ \frac{3 \epsilon}{M_Z^2} \frac{\sin^2\alpha\cos\alpha}{\sin \beta} \non \\
&& \lambda_{HHH} = 3\cos(2\alpha) \cos(\beta+\alpha)
+ \frac{3 \epsilon}{M_Z^2} \frac{\sin^3\alpha}{\sin \beta} \non \\
&& \lambda_{hAA} = \cos(2\beta) \sin(\beta+\alpha)
+ \frac{\epsilon}{M_Z^2} \frac{\cos\alpha\cos^2\beta}{\sin \beta} \non \\
&& \lambda_{HAA} = - \cos(2\beta) \cos(\beta+\alpha)
+ \frac{\epsilon}{M_Z^2} \frac{\sin\alpha\cos^2\beta}{\sin \beta}
\label{eq:4}
\end{eqnarray}
The couplings in the SM as well as the MSSM are normalized to
$\lambda_0 = [\sqrt{2} G_F]^{1/2} M_Z^2$. The MSSM couplings depend on
$\beta$ and the mixing angle $\alpha$
\begin{equation}
\tan 2 \alpha=
 \frac{ M_{A}^2 + M_Z^2 }{ M_{A}^2 - M_Z^2 + \epsilon / \cos 2\beta} 
 \; \tan 2 \beta
\end{equation}
in the CP--even Higgs sector; radiative corrections have been included
in the leading $m_t^4$ one--loop approximation, parametrized by
\begin{equation}
\epsilon = \frac{3 G_F}{\sqrt{2} \pi^2}
              \frac{m_t^4}{\sin^2 \beta} \;
              \log\left[ 1 + \frac{M_S^2}{m_t^2} \right]
\end{equation}
with the common squark mass fixed to $M_S$= 1 TeV.  In our numerical
analysis we have included the leading two--loop corrections to the
MSSM Higgs masses and couplings, taken from Ref.\cite{M1}. The
behaviour of the couplings with $M_A$ is shown for two representative
values of $\tb=1.5$ and 30 in Fig.1.

The paper is organized as follows. In the next section $HH$ pair
production through gluon collisions will be discussed in the Standard
Model.  In the third section we will present the results for all pair
combinations of neutral Higgs bosons in the MSSM. Phenomenological
aspects including Higgs--strahlung and $WW$ fusion will be presented
in a sequel to this paper \cite{8}.

\section[]{Higgs Pair Production in the Standard Model} 
Two mechanisms contribute to the production of Higgs pairs through
$gg$ collisions in the Standard Model, exemplified by the generic
diagrams in Fig.2a/b. (i) Virtual Higgs bosons which subsequently
decay into $HH$ final states, are coupled to gluons by the usual
heavy--quark triangle \cite{6,7}. (ii) The coupling is also mediated
by heavy--quark box diagrams.\smallskip

In the triangle diagram Fig.2a the gluons are coupled to the total
spin $S_z=0$ along the collision axis. The transition matrix element
associated with this mechanism can therefore be expressed by the
product of one (gauge invariant) form factor $F_\triangle$, depending
on the scaling variable $\tau_Q = 4 m_Q^2 /\hat{s}$, and the
generalized coupling $C_\triangle$ defined as
\begin{equation}     
  C_\triangle = \lambda_{HHH}~\frac{M_Z^2}{\hat s-M_H^2}
\end{equation}     
The coefficient $\lambda_{HHH}$ denotes the trilinear self--coupling
$HHH$ in the Standard Model, cf. eq.(\ref{eq:3}). $\hat{s}$ is the square of
the invariant energy flow through the virtual Higgs line. The
well--known form factor $F_\triangle$ \cite{6,7} is given in
Appendix~A1.

The box diagrams in Fig.2b allow for $S_z=0$ and 2 gluon--gluon
couplings so that the transition matrix element can be expressed in
terms of two (gauge invariant) form factors $F_\Box$ and $G_\Box$.
Fairly compact expressions of the form factors are given in
Appendix~A1 where the Higgs masses have to be specified to
$M_c=M_d=M_H$. The couplings between the Higgs bosons and the quarks
are normalized to unity by definition,
\begin{equation}
C_\Box = 1 
\end{equation}
[The overall normalization is included in the prefactors of the cross
section and the form factors.] The form factors obtained in this way
for the simplified case of equal masses agree with the results in
Ref.\cite{1}.

In the two limits of light and heavy Higgs bosons with respect to the
loop quark mass, very simple expressions can be derived for the three
form factors $F_\triangle$ and $F_\Box$, $G_\Box$.  \bigskip

\nn
\underline{\it (i) Large Quark Mass Limit} 
\bigskip

The form factors can be evaluated either by taking the limit $m_Q^2
\gg \hat{s} \sim M_H^2$ in the Feynman amplitude or, equivalently, by
exploiting the elegant low--energy theorem $F_\Box =
m^2_Q \partial (F_\triangle/m_Q)/\partial m_Q$
\citer{7,9} for external light scalar Higgs bosons.  This reduces the
complexity of the calculation considerably:
\begin{equation}
  F_\triangle=\frac{2}{3}+{\cal O}\left(\hat s/m_Q^2\right)
\label{eq:13}
\end{equation}
and 
\begin{eqnarray}
  F_\Box&=&-\frac{2}{3}+{\cal O}\left( {\hat s}/m_Q^2\right)
  \\ G_\Box&=&{\cal O}\left( {\hat s}/m_Q^2\right)
  \non
\end{eqnarray}
Both $S_z=0$ form factors $F_\triangle$ and $F_\Box$ survive in this
limit, whereas the $S_z=2$ form factor $G_\Box$ vanishes
asymptotically.  \bigskip

\nn
\underline{\it (ii) Small Quark Mass Limit} 
\bigskip

In the limit of light quark masses compared with the invariant
energy\footnote{In the leading logarithmic approximation the scale is
  set effectively by the Higgs mass since the invariant energy
  $\sqrt{\hat s}$ is of the same order as $M_H$ for the cross sections
  we shall study.}, large logarithms of the light quark mass $\log
\hat{s}/m_Q^2$ will occur in the form factors. Detailed inspection of
the Feynman amplitudes leads to the following final expressions:
\begin{equation}
  F_\triangle= - \frac{m_Q^2}{\hat s} \left[
  \log\left(\frac{m_Q^2}{\hat{s}} \right) + i\pi \right]^2 + {\cal
    O}\left(\frac{m_Q^2}{\hat s} \right) 
\end{equation}
and
\begin{eqnarray}
  F_\Box&=&{\cal O} \left( m_Q^2/{\hat s} \right) \label{eq:14}
  \\ G_\Box&=&{\cal O}\left( m_Q^2/{\hat s} \right) \non
\end{eqnarray}
All form factors vanish to ${\cal O}(m_Q^2)$ in this limit since they
are suppressed by the Yukawa coupling $\propto m_Q$; this is
well--known for the quark triangle. [This limit is not of much
practical use in the SM; yet it will be relevant later in the MSSM
where $b$--quark loops are dominant for large $\tb$.] \bigskip

The differential parton cross section can finally be written in the form
\begin{equation}  
\frac{d\hat\sigma (gg\to HH)}{d\hat t} = 
\frac{G_F^2\alpha_s^2}
{256 (2\pi)^3} \Big[ | C_\triangle \; F_\triangle \; + C_\Box F_\Box |^2 + |
C_\Box G_\Box |^2 \Big]
\label{eq:19}
\end{equation}  
where $\hat{t}$ is the momentum transfer squared from one of the
gluons in the initial state to one of the Higgs bosons in the final
state.  The total cross section for $HH$ Higgs pair production through
$gg$ in $pp$ collisions\footnote{In analogy to single Higgs production
  via gluon fusion, QCD corrections are expected to enhance the event
  rate significantly. However, they are known only for the triangle
  \cite{7} and not for the box; therefore they are not taken into
  account in the numerical analysis.} can be derived by integrating
(\ref{eq:19}) over the scattering angle and the $gg$ luminosity
\begin{equation}     
  \sigma(pp \to gg \to HH) = \int_{4 M_H^2 /s}^{1} d\tau \frac{d {\cal
      L}^{gg} }{d\tau} \hat{\sigma}(\hat s = \tau s)
\end{equation}     
This cross section has been evaluated numerically. The analysis has
been carried out for the LHC c.m. energy $\sqrt{s}$ = 14 TeV; the top
quark mass has been set to $m_t$ = 175 GeV \cite{10}. The result is
shown in Fig.3. For SM Higgs masses close to $M_H \sim$ 100 GeV the
cross section is of order 10 $fb$.  However, it drops rapidly with
increasing Higgs mass; this is a consequence of the fast fall--off of
the $gg$ luminosity with rising $\tau$ [which is of the order of
$x_g^2$ in the $pp$ collisions].  Nevertheless, at $M_H \sim$ 100 GeV
order 2,000 events will be produced for an integrated luminosity of 
$\int{\cal L} = 100 fb^{-1}$, giving rise to 4$b$ and
$bb\tau\tau$ final states with large transverse momenta.

\section{Neutral Higgs Pairs of the MSSM}
A large variety of neutral Higgs pairs of the MSSM can be produced in
gluon--gluon collisions:
\begin{equation}
  pp \to gg \to hh,hH,HH,hA,HA,AA
\end{equation}
The analysis of the cross sections is in general much more involved
than in the SM for two reasons. First, the masses of the two Higgs
bosons in the final state are in general different, and second,
besides pairs of CP--even Higgs bosons, also pairs of CP--odd and
mixed CP--even/CP--odd pairs can be produced.  \bigskip

\nn
{\bf a) Pairs of CP--even Higgs bosons $\mathbf{hh,hH,HH}$}
\bigskip

The mechanisms for the production of CP--even MSSM Higgs bosons $gg\to
H_i H_j$ $(H_i = h,H)$ are similar to the Standard Model. The generic
triangle and box diagrams are shown in Figs.4a/b. All possible
combinations of light and heavy Higgs bosons $h$ and $H$ can be
coupled in 3--particle vertices.

The matrix element of the triangle contributions can be expressed in
terms of the form factor $F_\triangle$, given in Appendix~A1, and the
couplings
\begin{equation}     
  C_\triangle = C_\triangle^h + C_\triangle^H
\end{equation}     
with
\begin{equation}     
  C^{h/H}_\triangle = \lambda_{H_iH_j(h/H)} ~\frac{M_Z^2}{\hat
    s-M_{h/H}^2+iM_{h/H}\Gamma_{h/H}}~g_Q^{h/H} \hspace{2cm} (H_i =
  h,H)
\end{equation}     
The couplings $g_Q^{h/H}$ denote the Higgs--quark couplings in units
of the SM Yukawa coupling $[\sqrt{2}G_F]^{1/2}m_Q$. They are collected
in Table \ref{tb:1}. The Higgs self couplings $\lambda_{H_i H_j
  (h/H)}$ can be read off eq.(\ref{eq:4}).
\begin{table}[hbt]
\renewcommand{\arraystretch}{1.5}
\begin{center}
\begin{tabular}{|lc||cc|} \hline
\multicolumn{2}{|c||}{$\Phi$} & $g^\Phi_t$
& $g^\Phi_b$ \\ \hline \hline
SM~ & $H$ & 1 & 1 \\ \hline
MSSM~ & $h$ & $\cos\alpha/\sin\beta$ & $-\sin\alpha/\cos\beta$ \\
& $H$ & $\sin\alpha/\sin\beta$ & $\cos\alpha/\cos\beta$ \\
& $A$ & $ 1/\tb$ & $\tb$ \\ \hline
\end{tabular} 
\renewcommand{\arraystretch}{1.2}
\caption[]{\label{tb:1}
{\it Higgs couplings in the MSSM to fermions relative to SM couplings.}}
\end{center}
\end{table}

For the box contributions we obtain analogous expressions, with two
independent form factors $F_\Box$ and $G_\Box$ contributing. Their
expressions can be found in Appendix~A1. The corresponding couplings
$C_\Box$ are are built up by the Higgs--quark couplings,
\begin{equation}     
C_\Box = g_Q^{H_i} g_Q^{H_j}
\end{equation}     

The two limiting cases in which the Higgs masses are either much
smaller or much larger than the loop quark masses, are both physically
relevant in this case since the top quark as well as the bottom quark
play a role, with $M^2_h \ll m_t^2$ and $M^2_H \gg m_b^2$. The limits
of the form factors are the same as for the SM Higgs bosons
eqs.(\ref{eq:13}--\ref{eq:14}).

From the differential parton cross section, which includes $t$ and $b$
quark loops,
\begin{equation}  
\frac{d\hat\sigma (gg\to H_iH_j)}{d\hat t} = 
\frac{G_F^2\alpha_s^2}{256 (2\pi)^3} 
\Big[ | \sum_{t,b} (C_\triangle \; F_\triangle \; 
       + C_\Box \; F_\Box) |^2
+ | \sum_{t,b} C_\Box \; G_\Box |^2 \Big]
\end{equation}  
the $pp$ cross section can be derived by integrating over the
scattering angle and attaching the gluon luminosity.  The result of
the numerical analysis is shown for the final states $hh,hH$ and $HH$
in Figs.5a/b/c for two representative values $\tb = 1.5$ and 30.
[Mixing in the stop sector is not taken into account.]

The final state $hh$ deserves special attention. The cross section for
moderate values $M_h$ is large since the heavy CP--even Higgs boson
$H$ can decay into the two light Higgs bosons (dashed lines). $H$ is
produced directly in $gg$ collisions, coupled through the quark
triangle, so that the matrix element involves one power of the
electroweak coupling less than continuum Higgs pair production.
Except for a dip due to a zero in the $Hhh$ coupling $\lambda_{Hhh}$,
the cross section in the resonance range is of order $10^3$ to $10^4
fb$, dropping to the typical $hh$ continuum value of 30 $fb$ for
increasing $H$ Higgs masses.  For $\tb=1.5$ the curve shows a sharp
threshold behavior for $M_H\sim 2 m_t$ due to on--shell top pair
production in the dominant top quark triangle and, moreover, due to
the sharp fall--off of the branching ratio $BR(H\to hh)$. The cross
section for $\tb=30$ depends strongly on the Higgs boson masses,
because resonance production $gg\to H \to hh$ is kinematically
forbidden for 110 GeV $\!\lsim M_A \lsim 210$ GeV.  Above this range
this channel opens up again with a small branching ratio $\sim
10^{-3}$ giving rise to the small bump in the cross section at
$M_A\sim 210$ GeV.

The cross sections for $hH$ and $HH$ final states are much smaller,
cf. Figs.5b/c.  This is the result of the absence of any resonance
effect and the phase space suppression due to the large Higgs masses
in the final states.  The cross section for $pp\to gg\to hH$ depends
strongly on $\tb$ and it varies between very small values and $\sim
100~fb$ if the entire Higgs mass range is sweeped. For $\tb=1.5$ the
Higgs mass dependence is smooth, whereas for $\tb=30$ a sharp peak
develops at $M_A\sim 100$ GeV due to the strong variation of the MSSM
couplings in this region. At large Higgs masses the cross section
decreases strongly with increasing $\tb$. A similar picture emerges
for the pair production of heavy scalar Higgs bosons $pp\to gg\to HH$.
Its size increases at large Higgs masses for increasing $\tb$, where
the bottom quark loops become dominant. The strong variation of the
cross section for $M_A\sim 100$ GeV and $\tb=30$ is again due to the
rapidly changing MSSM couplings in this mass range.  \bigskip

\nn
{\bf b) Mixed CP--even/CP--odd pairs}
\bigskip

The analysis of the CP--mixed $hA$ and $HA$ final states is in many
aspects different from the previous cases. First of all, the final
state can be produced through the decay of a virtual $Z$ boson as
shown in the generic diagrams of Fig.6.  By evaluating the Feynman
diagrams one finds the following generalized charges [$H_i = h,H$]
\begin{description}
\item[(i)~~] \underline{ $s$--channel $A$ exchange: }
\begin{equation}     
  C^A_\triangle = \lambda_{AAH_i}~\frac{M_Z^2}{\hat
    s-M_A^2+iM_A\Gamma_A}~g_Q^A 
\end{equation}
\item[(ii)~] \underline{ $s$--channel $Z$ exchange: }
\begin{equation}     
  C^Z_\triangle = \lambda_{ZAH_i}~\frac{M_Z^2}{\hat
    s-M_Z^2+iM_Z\Gamma_Z}~a_Q
\end{equation}
where $a_Q = 1 (-1)$ for top (bottom) quarks denotes the axial charge
of the heavy loop quark $Q$ [note that only the $Z$ axial coupling of
the heavy loop quark can contribute], and the trilinear couplings
$\lambda_{ZAH_i}$ are given by
\begin{eqnarray}
\lambda_{ZAh} & = & -\cos(\beta-\alpha) \\
\lambda_{ZAH} & = & \sin(\beta-\alpha) \non
\end{eqnarray}
\item[(iii)] \underline{ box contributions: }
\begin{equation}
C_\Box^{AH_i} = g_Q^A g_Q^{H_i}
\end{equation}
\end{description}
The expressions of the form factors $F_\triangle^{A/Z}$ and $F_\Box$,
$G_\Box$ can be found in Appendix~A2.

In analogy to the CP--even case, simple expressions can be derived for
the form factors in the limits of large and small Higgs masses
compared with the quark masses.  \bigskip

\nn
\underline{\it (i) Large Quark Mass Limit} 
\bigskip

The triangular form factor can be derived in this limit from the axial
anomaly. The box form factor is given by the derivative of the anomaly
$F_\Box = m_Q^2\partial(F_\triangle^A/m_Q)/\partial
m_Q$ \cite{7,9}. Simple expressions can be obtained:
\begin{eqnarray}
  F^A_\triangle&=&1+{\cal O}\left( {\hat s}/m_Q^2\right) \\ 
  F^Z_\triangle&=&{\cal O}\left( {\hat s}/m_Q^2\right) \non
\end{eqnarray}
and
\begin{eqnarray}
  F_\Box&=&-1+{\cal O}\left( {\hat s}/m_Q^2\right) \\ 
  G_\Box&=&{\cal O}\left( {\hat s}/m_Q^2\right) \non
\end{eqnarray}
\bigskip

\nn
\underline{\it (ii) Small Quark Mass Limit} 
\bigskip

\begin{eqnarray}
  F^A_\triangle&=& -\frac{m_Q^2}{\hat s}
  \left[\log\left(\frac{m_Q^2}{\hat{s}} \right) + i\pi \right]^2 +
    {\cal O}\left(\frac{m_Q^2}{\hat s} \right) \non \\ F^Z_\triangle
    &=& \frac{\hat s - M_Z^2}{\hat s}~\frac{M_{H_i}^2 - M_A^2}{M_Z^2}
    \left\{ 1 + \frac{m_Q^2}{\hat s}
      \left[\log\left(\frac{m_Q^2}{\hat{s}} \right) + i\pi \right]^2 +
        {\cal O}\left(\frac{m_Q^2}{\hat s} \right) \right\}\non \\ 
        F_\Box&=&{\cal O} \left( m_Q^2/{\hat s} \right) \\ 
        G_\Box&=&{\cal O}\left( m_Q^2/{\hat s} \right) \non
\end{eqnarray}
\bigskip

The differential parton cross section is determined by the form
factors, including $t$ and $b$ quark loops:
\begin{equation}  
\frac{d\hat\sigma (gg\to AH_i)}{d\hat t} = \frac{G_F^2\alpha_s^2}
{256 (2\pi)^3} 
\Big[ | \sum_{t,b} (C_\triangle \; F_\triangle \; 
       + C_\Box \; F_\Box) |^2
+ | \sum_{t,b} C_\Box \; G_\Box |^2 \Big]
\end{equation}  
The results of the numerical analysis of the two $pp$ cross sections
are shown in Figs.7a/b.
For light masses $M_h$ the invariant mass [$hA$] of the final state is
small so that the cross section is enhanced by the large phase space.
For large Higgs masses the cross section decreases strongly with
increasing $\tb$ and drops to a level of about $10^{-1}$ to $10^{-3}
fb$ at $M_A \sim 1$ TeV, where it cannot be observed anymore. The
strong variation for $\tb=30$ at $M_A\sim 100$ GeV is generated by the
rapid change of the MSSM couplings with $M_A$.

Due to the larger mass of the heavy scalar Higgs boson $H$, the cross
section for $HA$ production turns out to be smaller than for $hA$
production.  \bigskip

\nn
{\bf c) $\mathbf{AA}$ pairs of CP--odd Higgs bosons}
\bigskip

This case is again closely related to the CP--even pairs. In the box
diagrams, Fig.8b, some terms flip just the sign when the $\gamma_5
\times \gamma_5$ couplings are reduced to $\idnty$ in the loop trace.
As a result, the matrix elements can be expressed in terms of the
couplings
\begin{eqnarray}     
  C_\triangle & = & C^h_\triangle + C^H_\triangle
\end{eqnarray}     
with
\begin{eqnarray}     
  C^{h/H}_\triangle & = & \lambda_{AA(h/H)}~ \frac{M_Z^2}{\hat
    s-M_{h/H}^2+iM_{h/H}\Gamma_{h/H}}~g_Q^{h/H} \non \\ 
  C_\Box & = & (g_Q^A)^2
\end{eqnarray}
The form factors $F_\triangle$ for the triangle contributions and
$F_\Box$, $G_\Box$ for the box contribution can be found
in Appendix~A3 by choosing $M_c=M_d=M_A$.

In the limits of large and small Higgs masses compared with the quark
masses simple expressions can be derived.  \bigskip

\nn
\underline{\it (i) Large Quark Mass Limit } 
\bigskip

The triangle form factor coincides with the scalar expression
eq.(\ref{eq:13}), and the box form factor can be obtained from the
derivative of the gluon self--energy [${\cal M}(ggA^2) = \sqrt{2} G_F
m_Q \partial {\cal M}(gg)/\partial m_Q$] \cite{9} leading to the
results
\begin{eqnarray}
  F_\triangle&=&\frac{2}{3}+{\cal O}\left( {\hat s}/m_Q^2\right) 
\end{eqnarray}
and
\begin{eqnarray}
  F_\Box&=&\frac{2}{3}+{\cal O}\left( {\hat s}/m_Q^2\right) \\ 
  G_\Box&=&{\cal O}\left( {\hat s}/m_Q^2\right) \non
\end{eqnarray}
\bigskip

\nn
\underline{\it (ii) Small Quark Mass Limit } 
\medskip

\begin{equation}
F_\triangle= - \frac{m_Q^2}{\hat s} \left[ \log\left(\frac{m_Q^2}{\hat{s}}
               \right) + i\pi \right]^2
               +  {\cal O}\left(\frac{m_Q^2}{\hat s} \right) 
\end{equation}
and
\begin{eqnarray}
F_\Box&=&{\cal O} \left( m_Q^2/{\hat s} \right) \\
G_\Box&=&{\cal O}\left( m_Q^2/{\hat s} \right) \non
\end{eqnarray}
\bigskip

The $pp$ cross section is shown in Fig.9. The large value of the cross
section for small $A$ masses is due to resonance $H\to AA$ decays.
When this channel is closed, the cross section becomes more and more
suppressed with rising $A$ mass in the final state.  For $\tb=1.5$ the
cross section develops a kink at the $(t\bar t)$ threshold $M_A=2m_t$
due to $S$--wave $(t\bar t)$--resonance production.  Above this mass
value the phase space suppression leads to a rapidly decreasing cross
section, to a level of $\sim 10^{-3} fb$ at $M_A\sim 1$ TeV. For
$\tb=30$ the bottom quark loops are dominant so that the cross section
depends smoothly on $M_A$. For large pseudoscalar masses the signal
increases with increasing $\tb$.


\section{Appendices}
\nn
{\bf A1. Form factors for two scalar Higgs bosons} 
         $\mathbf{ g_a g_b \to H_c H_d}$
\medskip

\nn
\underline{\it notation:}
\medskip 

\nn
parameter definitions:
\begin{displaymath}
\hat s = (p_a+p_b)^2, \hspace{1.0cm}
\hat t = (p_c-p_a)^2, \hspace{1.0cm}
\hat u = (p_c-p_b)^2
\end{displaymath}
\begin{displaymath}
S = {\hat s}/m_Q^2, \hspace{1.0cm}
T = {\hat t}/m_Q^2, \hspace{1.0cm}
U = {\hat u}/m_Q^2
\end{displaymath}
\begin{displaymath}
\rho_c = M_c^2/m_Q^2, \hspace{1.0cm}
\rho_d = M_d^2/m_Q^2, \hspace{1.0cm}
\tau_Q = 4/S
\end{displaymath}
\begin{displaymath}
T_1 = T - \rho_c, \hspace{1.0cm}
U_1 = U - \rho_c, \hspace{1.0cm}
T_2 = T - \rho_d, \hspace{1.0cm}
U_2 = U - \rho_d
\end{displaymath}

\nn
Scalar integrals:
\begin{eqnarray*}
C_{ij}\!\!\!\! & = & \!\!\!\!\!\int \frac{d^4q}{i\pi^2}~\frac{1}
{(q^2-m_Q^2)\left[\pht (q+p_i)^2-m_Q^2\right]
\left[\pht (q+p_i+p_j)^2-m_Q^2\right]} \\ \\
D_{ijk}\!\!\!\! & = & \!\!\!\!\!\int \frac{d^4q}{i\pi^2} \frac{1}
{(q^2-m_Q^2)\left[\pht (q+p_i)^2-m_Q^2\right]
\left[\pht (q+p_i+p_j)^2-m_Q^2\right]\left[\pht (q+p_i+p_j+p_k)^2-m_Q^2\right]}
\end{eqnarray*}
The analytic expressions can be found in Ref.\cite{7b}.

\bigskip 

\nn
\underline{\it triangle form factor:}
\begin{eqnarray*}
F_\triangle & = & \frac{2}{S} \left\{ 2+(4 - S) m_Q^2 C_{ab} \right\}
= \tau_Q \left[\pht 1+(1-\tau_Q) f(\tau_Q)\right]
\end{eqnarray*}

\begin{eqnarray*}
f(\tau_Q)=\left\{
\begin{array}{ll}  \displaystyle
\arcsin^2\frac{1}{\sqrt{\tau_Q}} & \tau_Q \geq 1 \\
\displaystyle -\frac{1}{4}\left[ \log\frac{1+\sqrt{1-\tau_Q}}
{1-\sqrt{1-\tau_Q}}-i\pi \right]^2 \hspace{0.5cm} & \tau_Q<1
\end{array} \right.
\end{eqnarray*}

\nn
\underline{\it box form factors:}
\begin{eqnarray*}
F_\Box & = & \frac{1}{S^2}\left\{ \pht 4 S +  8 S m_Q^2 C_{ab}
-2S(S+\rho_c+\rho_d-8)m_Q^4 (D_{abc}+D_{bac}+D_{acb}) \right. \\
& + & \left. (\rho_c+\rho_d-8)m_Q^2\!\left[ \pht T_1 C_{ac} + U_1 C_{bc}
+ U_2 C_{ad} + T_2 C_{bd} - (TU-\rho_c\rho_d) m_Q^2 D_{acb}\right] \right\} \\
G_\Box & = & \frac{1}{S(TU-\rho_c\rho_d)}\left\{ \pht (T^2+\rho_c\rho_d-8T)
m_Q^2 \left[ \pht S C_{ab} + T_1 C_{ac} +T_2 C_{bd} - ST m_Q^2 D_{bac} \right]
\right. \\
& & \hspace{2.55cm}+(U^2+\rho_c\rho_d-8U)m_Q^2\left[ \pht S C_{ab} + U_1 C_{bc}
+U_2 C_{ad} - SU m_Q^2 D_{abc} \right] \\
& & \hspace{2.55cm} - (T^2+U^2-2\rho_c\rho_d) (T+U-8) m_Q^2 C_{cd} \\
& & \hspace{2.55cm} \left. - 2(T+U-8)(TU-\rho_c\rho_d) m_Q^4
(D_{abc}+D_{bac}+D_{acb}) \pht \right\}
\end{eqnarray*}

\nn
\underline{\it tensor basis:}
\begin{eqnarray*}
S_z=0&:&\qquad A_1^{\mu\nu} = g^{\mu\nu}-\frac{p_a^\nu p_b^\mu }{(p_a p_b)} \\
S_z=2&:&\qquad A_2^{\mu \nu} = g^{\mu \nu}
               +\frac{p_c^2 p_a^\nu p_b^\mu}{p_T^2 (p_a p_b)}
               -\frac{2 (p_b p_c) p_a^\nu p_c^\mu}{p_T^2 (p_a p_b)}
               -\frac{2 (p_a p_c) p_b^\mu p_c^\nu}{p_T^2 (p_a p_b)}
               +\frac{2 p_c^\mu p_c^\nu}{p_T^2} \\ \\
&&\text{with} \qquad \qquad p_T^2 =
2\frac{(p_ap_c)(p_bp_c)}{(p_ap_b)}-p_c^2 \\ \\
&&A_1 \cdot A_2 = 0 , \qquad A_1 \cdot A_1 = A_2 \cdot A_2 = 2
\end{eqnarray*}

\nn
\underline{\it matrix element:}
\begin{eqnarray*}     
{\cal M}\left(g_ag_b \to H_cH_d \right)&=&
        {\cal M}_\triangle^h
       +{\cal M}_\triangle^H
       +{\cal M}_\Box \\
{\cal M}_\triangle^{h/H}&=& 
        \frac{G_F \alpha_s \s }{2 \sqrt{2} \pi} \;
        C_\triangle^{h/H} \; F_\triangle {A_1}_{\mu \nu} \; 
        \epsilon_a^\mu \epsilon_b^\nu \; \delta_{ab} \\
{\cal M}_\Box&=& 
        \frac{G_F \alpha_s \s }{2 \sqrt{2} \pi} \;
        C_\Box  \; ( F_\Box {A_1}_{\mu \nu} + G_\Box {A_2}_{\mu \nu} ) \; 
        \epsilon_a^\mu \epsilon_b^\nu \; \delta_{ab} 
\end{eqnarray*}     

\nn
{\bf A2. Form factors for a mixed scalar--pseudoscalar pair} 
         $\mathbf{g_a g_b \to A_c H_d}$
\bigskip

\nn
\underline{\it triangle form factor:}
\begin{eqnarray*}
F_\triangle^A & = & -2 \mq C_{ab} = \tau_Q f(\tau_Q) \\
F_\triangle^Z & = & \left( 1-\frac{\s}{M_Z^2} \right) \frac{\rho_c-\rho_d}{S}
\left[ \pht 1 + 2 m_Q^2 C_{ab} \right]
= \left( 1-\frac{\s}{M_Z^2} \right) \frac{\rho_c - \rho_d}{S}
\left[ \pht 1 - \tau_Q f(\tau_Q) \right]
\end{eqnarray*}
\medskip 

\nn
\underline{\it box form factors:}
\begin{eqnarray*}
F_\Box & = & \frac{1}{S^2}\left\{ \pht
-2S(S+\rho_c-\rho_d) m_Q^4 (D_{abc}+D_{bac}+D_{acb}) \right. \\
& & \left. + (\rho_c-\rho_d)m_Q^2 \left[ \pht T_1 C_{ac} + U_1 C_{bc}
+ U_2 C_{ad} + T_2 C_{bd} - (TU-\rho_c\rho_d) m_Q^2 D_{acb}\right] \right\} \\
G_\Box & = & \frac{1}{S(TU-\rho_c\rho_d)}\left\{ \pht (U^2-\rho_c\rho_d)
m_Q^2 \left[ \pht S C_{ab} + U_1 C_{bc} + U_2 C_{ad} - SU m_Q^2 D_{abc} \right]
\right. \\
& & \hspace{2.60cm} -(T^2-\rho_c\rho_d) m_Q^2
\left[ \pht S C_{ab} + T_1 C_{ac} + T_2 C_{bd} - ST m_Q^2 D_{bac} \right] \\
& & \hspace{2.60cm} +\left[\pht (T+U)^2-4\rho_c\rho_d\right](T-U)m_Q^2C_{cd} \\
& & \hspace{2.60cm} \left. + 2(T-U)(TU-\rho_c\rho_d) m_Q^4
(D_{abc}+D_{bac}+D_{acb}) \pht \right\}
\end{eqnarray*}

\nn
\underline{\it tensor basis:}
\begin{eqnarray*}
S_z=0&:&\qquad A_1^{\mu\nu} = \frac{1}{(p_a p_b)} \epsilon^{\mu \nu p_a p_b} \\
S_z=2&:&\qquad A_2^{\mu \nu} = \frac{ p_c^\mu \epsilon^{\nu p_a p_b p_c}
               +p_c^\nu \epsilon^{\mu p_a p_b p_c}
               +(p_b p_c) \epsilon^{\mu \nu p_a p_c}
               +(p_a p_c) \epsilon^{\mu \nu p_b p_c} }
               {(p_a p_b) p_T^2} \\ \\
&&\text{with} \qquad \qquad p_T^2 =
2\frac{(p_ap_c)(p_bp_c)}{(p_ap_b)}-p_c^2 \\ \\
&&A_1 \cdot A_2 = 0 \qquad A_1 \cdot A_1 = A_2 \cdot A_2 = 2
\end{eqnarray*}

\nn
\underline{\it matrix element:}
\begin{eqnarray*}     
{\cal M}\left(g_ag_b \to A_cH_d \right)&=&
        {\cal M}_\triangle^A
       +{\cal M}_\triangle^Z
       +{\cal M}_\Box \\
{\cal M}_\triangle^{A/Z}&=& 
        \frac{G_F \alpha_s \s }{2 \sqrt{2} \pi} \;
        C_\triangle^{A/Z} \; F_\triangle^{A/Z} {A_1}_{\mu \nu} \; 
        \epsilon_a^\mu \epsilon_b^\nu \; \delta_{ab} \\
{\cal M}_\Box&=& 
        \frac{G_F \alpha_s \s }{2 \sqrt{2} \pi} \;
        C_\Box \; ( F_\Box {A_1}_{\mu \nu} + G_\Box {A_2}_{\mu \nu} ) \;
        \epsilon_a^\mu \epsilon_b^\nu \; \delta_{ab}   
\end{eqnarray*}     
\bigskip 

\nn {\bf A3. Form factors for two pseudoscalar Higgs
  bosons}\footnote{We present the formulae for different masses of the
  two pseudoscalar particles so that they can also be applied to
  non--minimal SUSY extensions in which several pseudoscalar particles
  in general belong to the Higgs spectrum.}
         $\mathbf{ g_a g_b \to A_c A_d}$
\bigskip

\nn
\underline{\it triangle form factor:}
\begin{eqnarray*}
F_\triangle & = & \frac{2}{S} \left\{ 2+(4 - S) m_Q^2 C_{ab} \right\}
= \tau_Q \left[\pht 1+(1-\tau_Q) f(\tau_Q)\right]
\end{eqnarray*}

\nn
\underline{\it box form factors:}
\begin{eqnarray*}
\AA
\end{eqnarray*}
 
\nn
\underline{\it tensor basis:}
\begin{eqnarray*}
S_z=0&:&\qquad A_1^{\mu\nu} = g^{\mu\nu}-\frac{p_a^\nu p_b^\mu }{(p_a p_b)} \\
S_z=2&:&\qquad A_2^{\mu \nu} = g^{\mu \nu}
               +\frac{p_c^2 p_a^\nu p_b^\mu}{p_T^2 (p_a p_b)}
               -\frac{2 (p_b p_c) p_a^\nu p_c^\mu}{p_T^2 (p_a p_b)}
               -\frac{2 (p_a p_c) p_b^\mu p_c^\nu}{p_T^2 (p_a p_b)}
               +\frac{2 p_c^\mu p_c^\nu}{p_T^2} \\ \\
&&\text{with} \qquad \qquad p_T^2 =
2\frac{(p_ap_c)(p_bp_c)}{(p_ap_b)}-p_c^2 \\ \\
&&A_1 \cdot A_2 = 0 \qquad A_1 \cdot A_1 = A_2 \cdot A_2 = 2
\end{eqnarray*}

\nn
\underline{\it matrix element:}
\begin{eqnarray*}     
{\cal M}\left(g_ag_b \to A_cA_d \right)&=&
        {\cal M}_\triangle^h
       +{\cal M}_\triangle^H
       +{\cal M}_\Box \\
{\cal M}_\triangle^{h/H}&=& 
        \frac{G_F \alpha_s \s }{2 \sqrt{2} \pi} \;
        C_\triangle^{h/H}  \; F_\triangle {A_1}_{\mu \nu} \; 
        \epsilon_a^\mu \epsilon_b^\nu \; \delta_{ab} \\
{\cal M}_\Box&=& 
        \frac{G_F \alpha_s \s }{2 \sqrt{2} \pi} \;
        C_\Box \;  ( F_\Box {A_1}_{\mu \nu} + G_\Box {A_2}_{\mu \nu} ) \; 
        \epsilon_a^\mu \epsilon_b^\nu \; \delta_{ab}
\end{eqnarray*}

\newpage

\nn
{\large \bf FIGURE CAPTIONS}
\\

\nn {\bf Fig.1:} Trilinear Higgs couplings in the MSSM as a function 
of the pseudoscalar Higgs mass $M_A$ for two representative values 
$\tb=1.5$ and 30 [no mixing in the stop sector, c.f. Ref.\cite{M1}].
\\

\nn {\bf Fig.2:} Generic diagrams describing SM Higgs pair production
in gluon--gluon collisions: a) triangle and b) box contributions.
\\

\nn {\bf Fig.3:} Total cross section of SM Higgs pair production at
the LHC [$\sqrt{s} = 14$ TeV]. The top mass has been chosen to be $m_t
= 175$ GeV and the bottom mass $m_b = 5$ GeV. The GRV parametrizations
\cite{11} of the parton densities have been adopted. The factorization
and renormalization scales are identified with the invariant mass of
the Higgs boson pair. The world average value $\alpha_s(M_Z) = 0.118$
\cite{12} has been chosen for the QCD coupling.
\\

\nn {\bf Fig.4:} Generic diagrams contributing to pair production of
CP--even MSSM Higgs bosons in gluon--gluon collisions $gg\to hh,
hH,HH$: a) triangle and b) box contributions.
\\

\nn {\bf Fig.5:} Total cross sections for pair production of CP--even
MSSM Higgs bosons in gluon--gluon collisions at the LHC: a) $hh$, b)
$hH$ and c) $HH$ pair production for $\tb=1.5$ and 30. The secondary
axes present the scalar Higgs masses $M_{h/H}$ corresponding to the
pseudoscalar masses $M_A$ for both values of $\tb$. The same
parameters and parton densities as in Fig.3 have been adopted.
The dashed curves in Fig.5a represent the resonant contributions
$gg\to H\to hh$.
\\

\nn {\bf Fig.6:} Diagram of the $Z$ boson $s$--channel contribution to
mixed CP--even/CP--odd MSSM Higgs pair production in gluon--gluon
collisions and box diagram.
\\

\nn {\bf Fig.7:} Total cross sections for mixed CP--even/CP--odd pair
production of MSSM Higgs bosons in gluon--gluon collisions at the LHC
for $\tb=1.5$ and 30: a) $hA$ and b) $HA$ production. The secondary
axes show the scalar Higgs boson masses corresponding to the
pseudoscalar mass $M_A$ for both values of $\tb$. The same parameters
as in Fig.3 have been chosen.
\\

\nn {\bf Fig.8:} Generic triangle and box diagrams for pair production
of CP--odd MSSM Higgs bosons in gluon--gluon collisions.
\\

\nn {\bf Fig.9:} Total cross section of CP--odd MSSM Higgs pair
production in gluon--gluon collisions at the LHC for $\tb=1.5$ and 30.
The parameters are the same as in Fig.3.
The dashed curves represent the resonant contributions $gg\to H\to AA$.

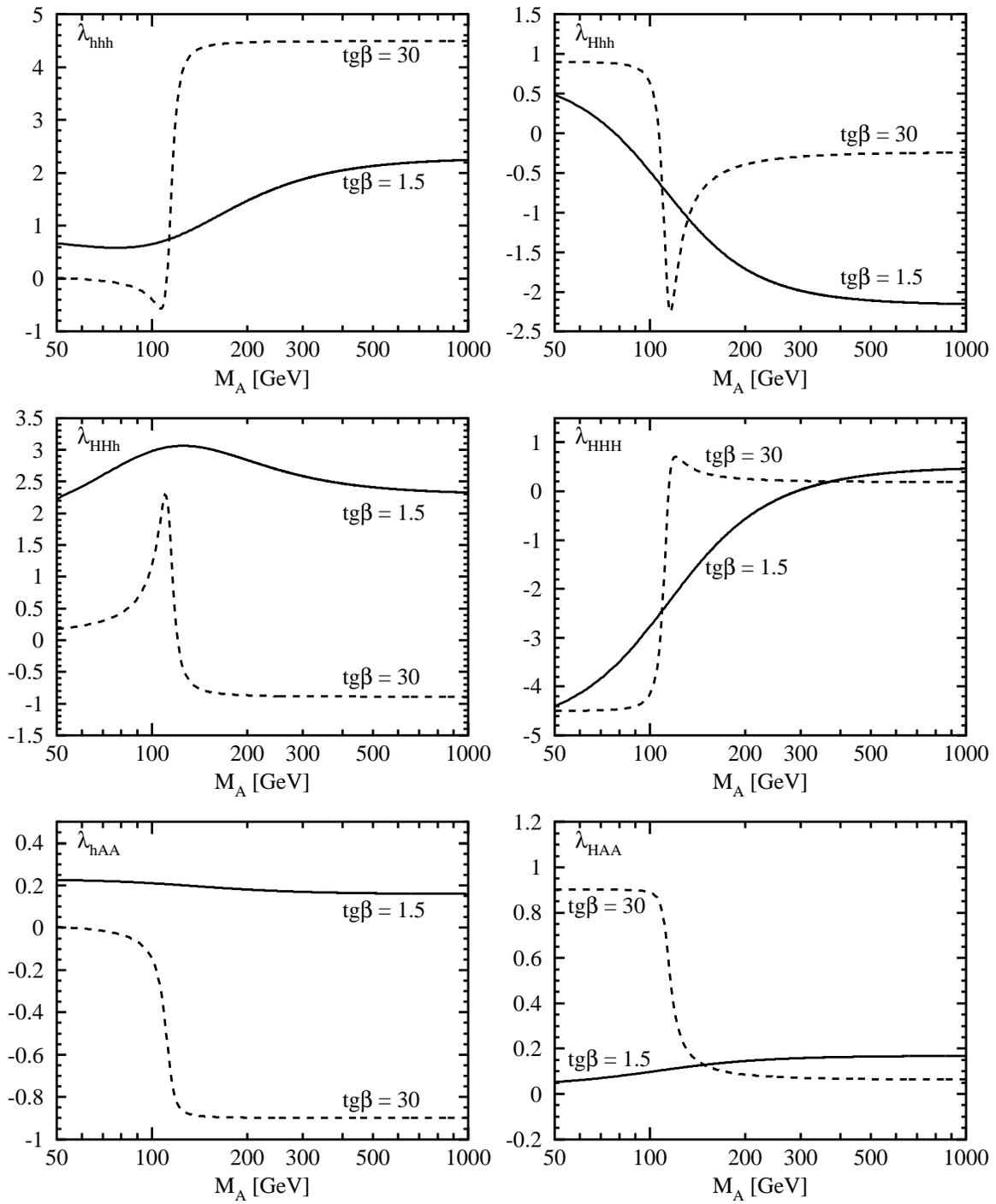
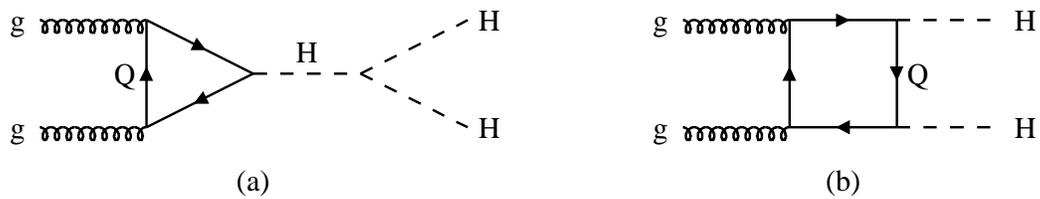
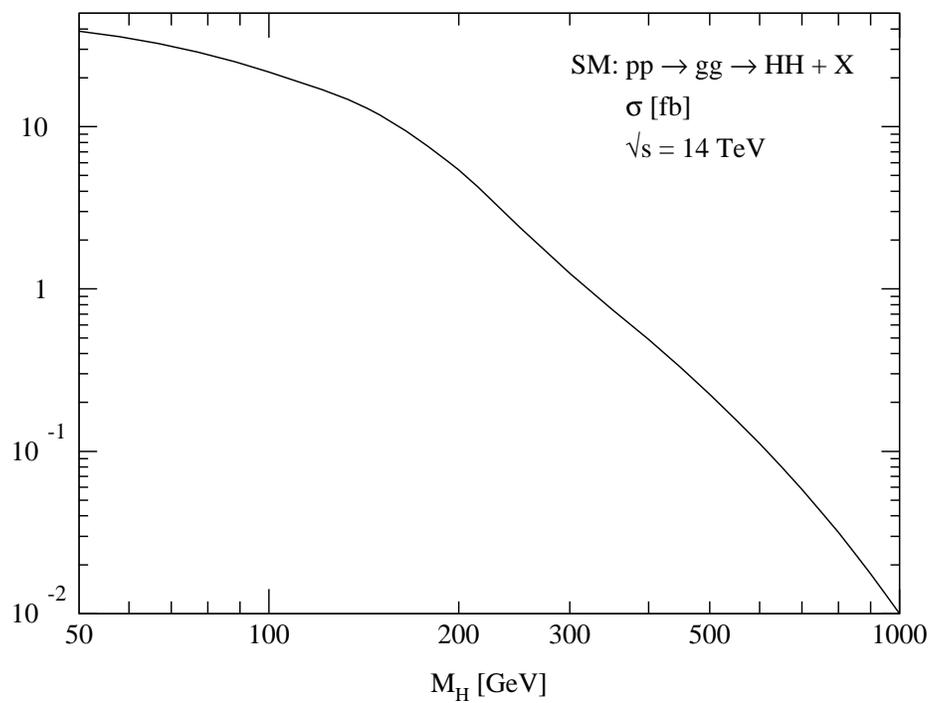
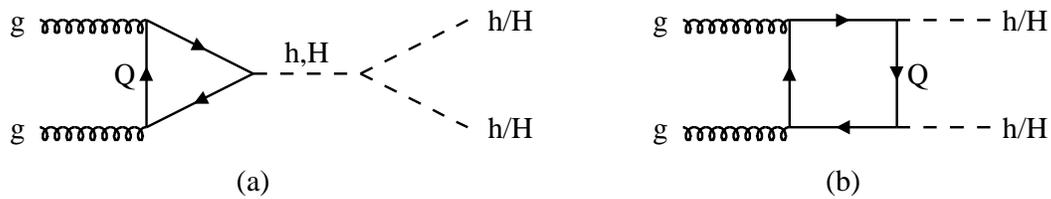
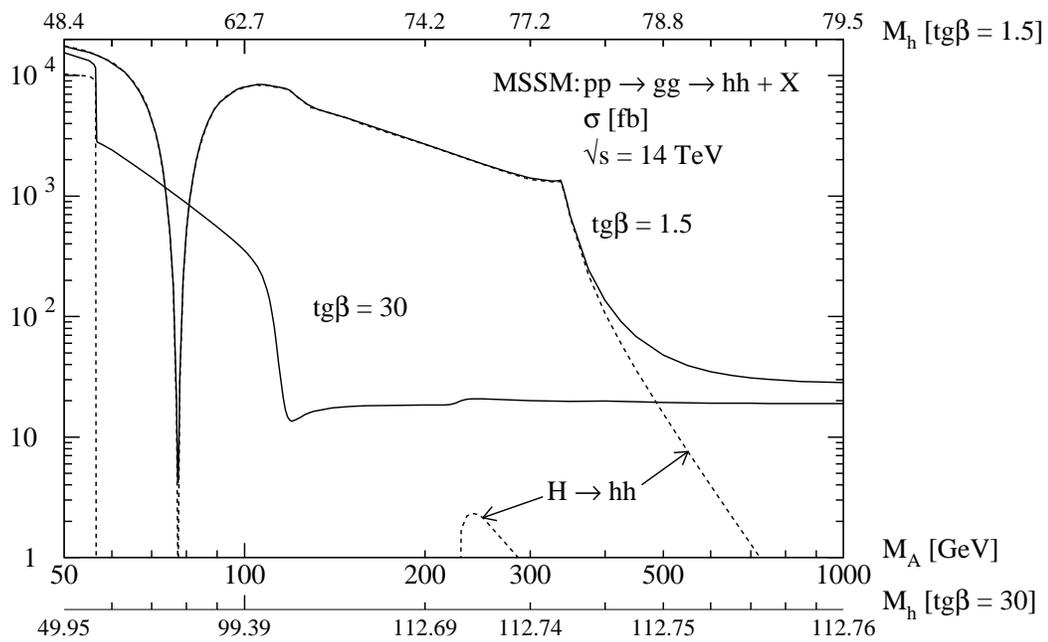
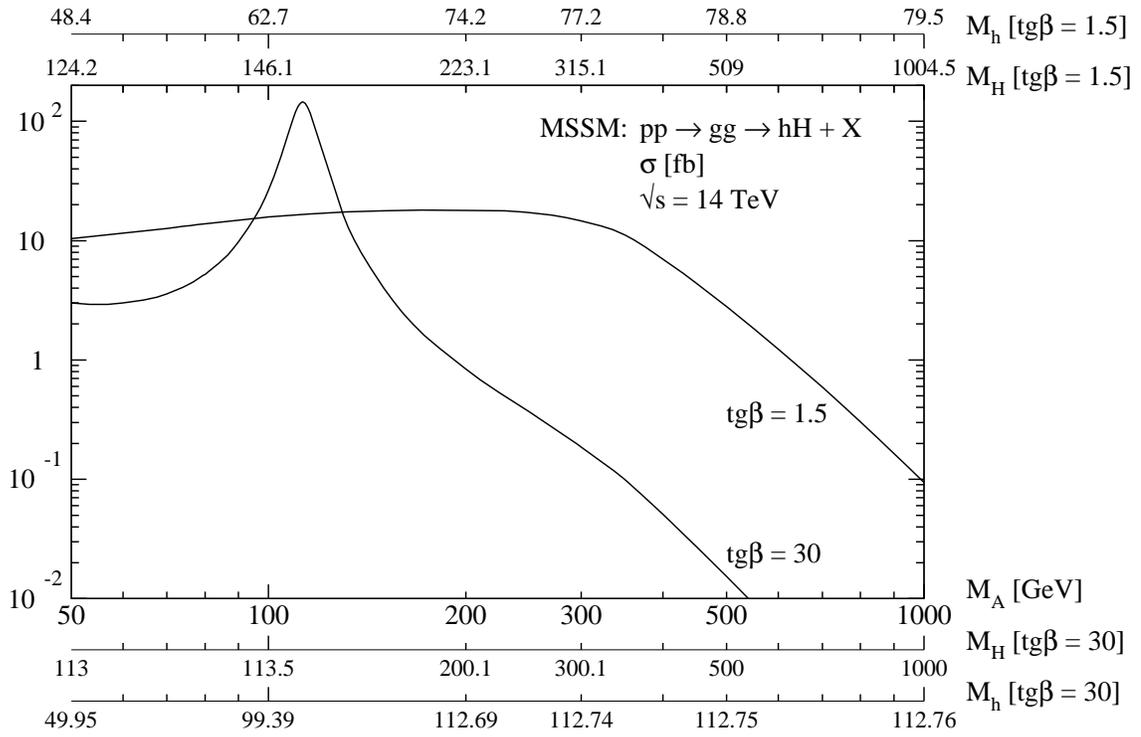
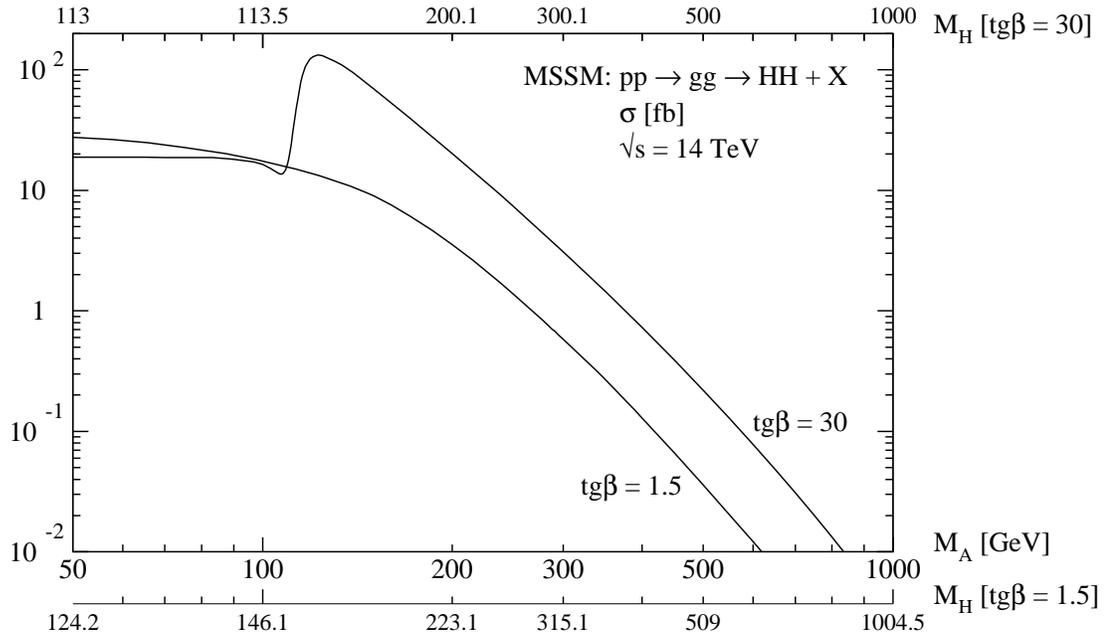
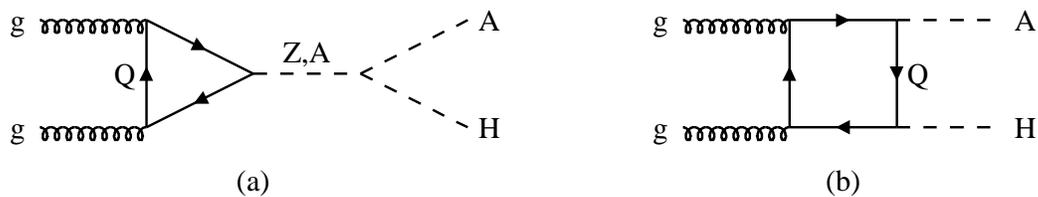
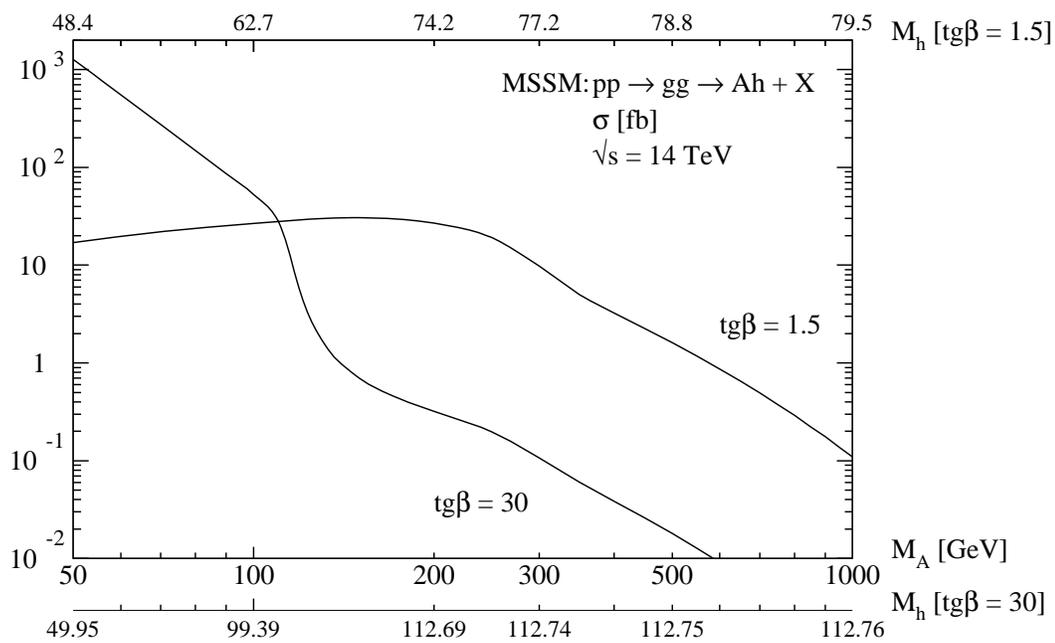
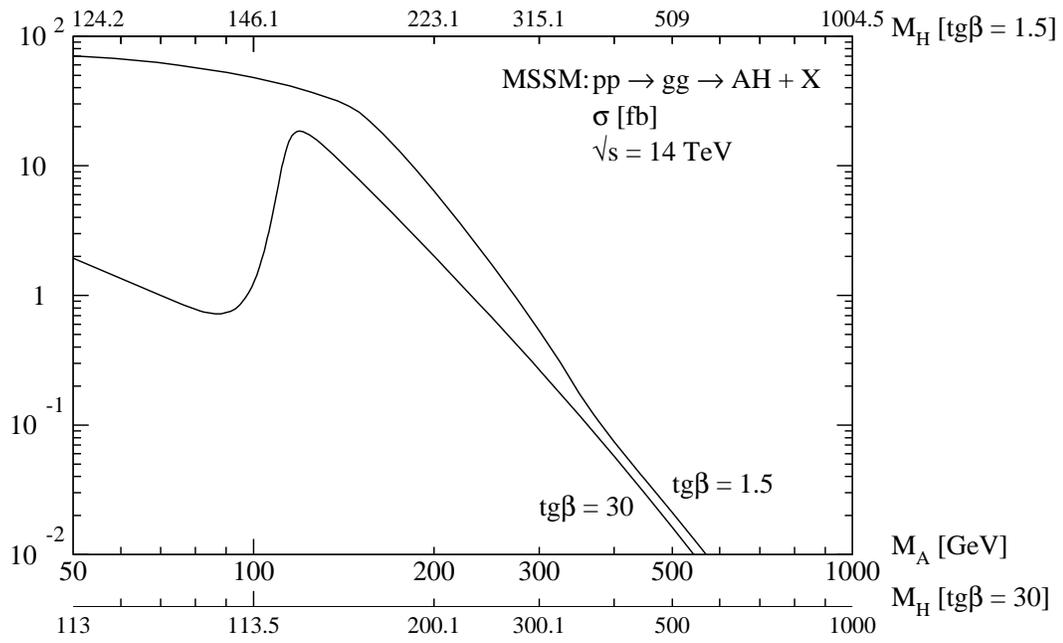
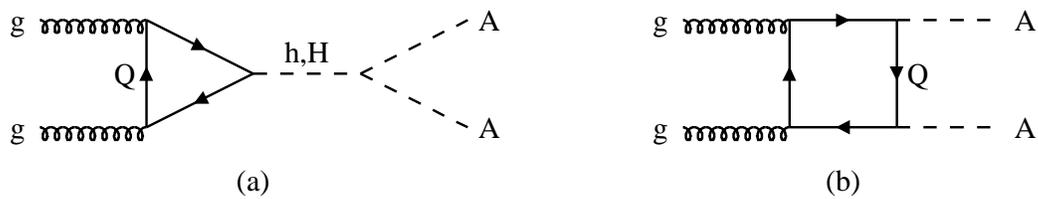
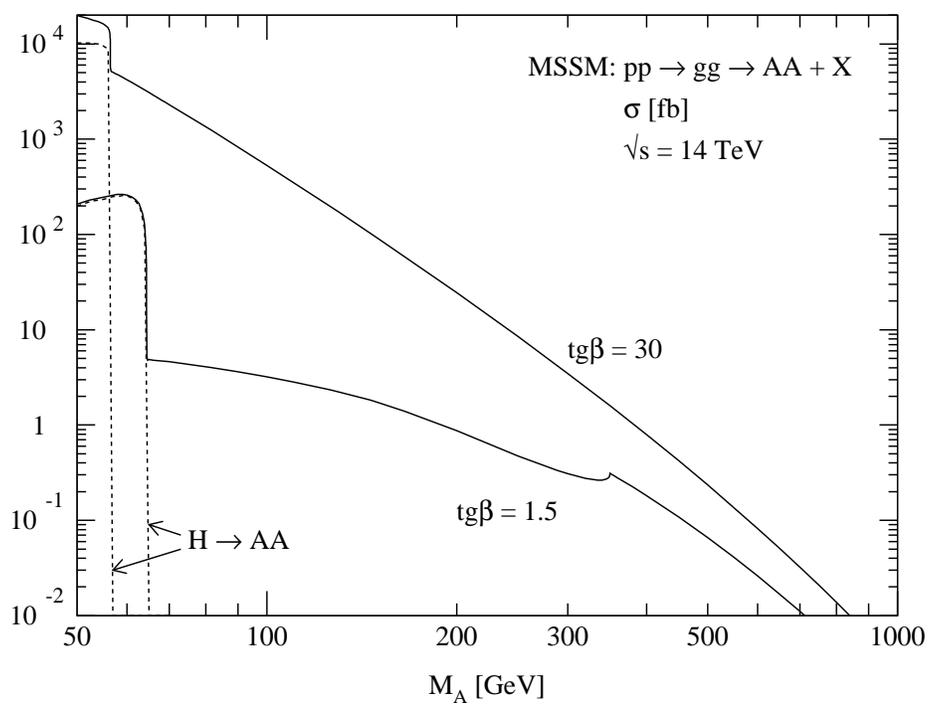
\end{document}